\documentclass[a4paper]{article}

\usepackage{graphicx}
\usepackage{dcolumn}
\usepackage{latexsym}
\usepackage{bm}

\begin{document}
\title{The condensate for two dynamical chirally improved quarks in QCD}
\author{C. B. Lang \\
Institut f\"ur Physik, FB Theoretische Physik,
Universit\"at Graz, Austria 
\footnote{christian.lang@uni-graz.at} \and
Pushan Majumdar \\
Institut f\"ur Theoretische Physik,
Westf\"alische Wilhelms Universit\"at M\"unster, Germany 
\footnote{pushan@uni-muenster.de} \and
Wolfgang Ortner \\
Institut f\"ur Physik, FB Theoretische Physik,
Universit\"at Graz, Austria 
\footnote{wolfgang.ortner@uni-graz.at} \and
(Bern-Graz-Regensburg (BGR)
collaboration)}

\maketitle

\begin{abstract}
We compare the eigenvalue spectra of the Dirac operator from a simulation
with two mass degenerate dynamical chirally  improved fermions with Random
Matrix Theory. Comparisons with distribution of $k-$th eigenvalues
$(k=1,2)$ in fixed topological sectors $(\nu=0,1)$ are carried out using
the Kolmogorov-Smirnov test. The  eigenvalue distributions are well
described by the RMT  predictions. The match allows us to read off the quark
condensate in the  chiral limit. Correcting for finite size and
renormalization we obtain a mean value of $-(276 (11)(16)\textrm{~MeV})^3$
in the \ensuremath{\overline{\textrm{MS}}}  scheme.
\end{abstract}


\section{Introduction}

The light quark condensate 
$\Sigma = \langle \bar u u\rangle \approx\langle \bar d d\rangle $ 
is a measure of chiral symmetry breaking. Most prominently it is the
proportionality factor  between the renormalized  quark mass and the
experimentally measurable $(f_\pi M_\pi)^2$ (the Gell-Mann--Oakes--Renner
relation \cite{GeOaRe68}) in the chiral limit. In the effective description
of low energy chiral symmetry breaking (chiral perturbation theory
\cite{We79GaLe84}) it is a fundamental externally supplied parameter. Other
features, like the microscopic eigenvalue distribution of the Dirac
operator, follow universal laws related just to the general symmetry
structure -- there the condensate is {\em the} only parameter which is to be
provided. Only full QCD calculations include the necessary dynamics to
determine this parameter ab initio.

QCD breaks chiral symmetry spontaneously, the small quark masses provide an
additional explicit breaking. The condensate is affected by both, but as far
as is known, only weakly by the explicit breaking, i.e., it has a
distinctive non-zero value in the chiral limit. Both, the physical quark
mass and the condensate are renormalization prescription dependent and have
to be given in some scheme. In the continuum
\ensuremath{\overline{\textrm{MS}}}  scheme (at scale 2 GeV) the condensate
has been determined from QCD sum rules to values between  
$\Sigma^{1/3}=-250\ldots -270 \textrm{~MeV}$ \cite{Na89Ja02}.
A recent large-N expansion gives similar values, not very sensitive to 
the number of flavors \cite{ArShVe0406}.
 
Lattice methods to compute $\Sigma$ include measurements of ratios of
correlation functions (for recent determinations in the framework of the
Dirac operator used here, the Chirally Improved (CI) operator, see
\cite{GaHuLa05a}). These, however, need lattices of sufficient size to
identify the asymptotic behavior of the correlators and from these the
renormalized quark mass. Random matrix theory, on the other hand, allows the
determination on small lattices.

Older global averages, using two dynamical flavors of   staggered, Wilson
and clover improvement type of  fermions generated values like
$\Sigma^{1/3}=-280(13) \textrm{~MeV}$ for the chiral condensate
\cite{GuBh97}.  More recent determinations using parameters of chiral
Lagrangians obtained from lattice  QCD \cite{Mc05} yielded values like
$\Sigma^{1/3}=-259(27) \textrm{~MeV}$ in the
\ensuremath{\overline{\textrm{MS}}}  at $(2\textrm{~GeV})$ for 2+1 flavors
of dynamical  staggered quarks (cf. that reference also for a comparison
with other authors' values). Domain wall fermions have also been used to
measure the condensate for two dynamical flavors \cite{AoBlCh05}, giving
values of the  unrenormalized $\Sigma^{1/3}$ between $-223$ MeV and $-261$
MeV.

Studies with dynamical overlap fermions on small ($10^4$) lattices, using
RMT distribution like we do here, report values of $\Sigma^{1/3}=-269(9)
\textrm{~MeV}$ for  $N_f=1$ \cite{DeHoLi06} and $\Sigma^{1/3}=-282(10)
\textrm{~MeV}$ for $N_f=2$ \cite{DeLiSc06}.  The latter values are without a
finite volume (lattice shape) correction factor, which would lower the value
to $-269$ MeV.\footnote{After completion of this paper we learned
of recent work \cite{JLQCD} presenting a two-flavor simulation with
overlap quarks in the $\epsilon$-regime. The result quoted there is
-251(7)(11) MeV.}

A recent study with staggered sea quarks and overlap valence quarks obtain 
an unrenormalized   value of $-291(5)$ MeV for $\Sigma^{1/3}$
\cite{HaHo06}, with estimated renormalization and finite volume 
corrections to be ~13 -- 18\%. Our own studies with 2 dynamical chirally
improved fermions, based on the GMOR relation,  indicated a value of 
$\Sigma^{1/3}=-288 (8) \textrm{~MeV}$ \cite{LaMaOr05c}.

Here we study spectra of the CI Dirac operators, obtained in our dynamical
simulation, comparing the resulting density distributions with the RMT
parameterizations in order to obtain values for the condensate including
finite size corrections. The approach is similar to that in \cite{DeLiSc06},
the difference being mainly a different Dirac operator and larger lattices,
here at four different combinations of quark mass and lattice spacing.

\section{Technical background}
\subsection{Simulation with CI fermions}

The CI Dirac operator is an approximate Ginsparg-Wilson operator
\cite{GiWi82}; it is a truncated
series solution  \cite{Ga01aGaHiLa00} to the GW relation. CI fermions have been already
extensively tested in {\em quenched} calculations (see, e.g.,
Ref.~\cite{GaGoHa03a}). There it was found that one can reach pion masses
below 300 MeV without running into the problem of exceptional configurations
(spurious zero modes).  On quenched  configurations pion masses down to
280~MeV could be obtained on lattices of size \(16^3 \times 32\) (lattice
spacing 0.148~fm).

In recent works~\cite{LaMaOr05ab,LaMaOr05c,LaMaOr06} we have
studied  the CI fermions in a dynamical simulation of QCD with two  light
flavors. All technicalities are discussed in Ref.~\cite{LaMaOr05c}. We use
the  L\"uscher-Weisz gauge action~\cite{LuWe85} and stout
smearing~\cite{MoPe04} of the gauge fields as part of the Dirac operator
definition. The Hybrid Monte Carlo method was implemented to deal with the
dynamics of the fermions. The lattices were (up to now) of moderate size:
$8^3\times 16$ and $12^3\times 24$, with lattice spacings between 0.11 and
0.14 fm. Table \ref{tab:runparameters} summarizes the simulation parameters
of the runs discussed here (see, however, Ref.~\cite{LaMaOr05c} for a more
complete list and details of the simulation).

\begin{table}[t]
\begin{center}
\begin{tabular}{c|rrrrrrrrr}
\hline
run & \(L^3 \times T\)  & $\beta_1$  & \(a\,m\)   &$^{\rm HMC}_{\rm ~time}$  &\(a_S\)[fm]    & \(a\,m_{AWI}\)     &conf& $\rho$&$Z_{S,q}$\\
\hline
a&    \(12^3 \times 24\)& 5.2        &  0.02      &463        &  0.115(6)   &  0.025     & 73  & 1.21  & 1.07  \\
b&    \(12^3 \times 24\)& 5.2        &  0.03      &363        &  0.125(6)   &  0.037    & 52  & 1.18  & 1.10  \\
c&    \(12^3 \times 24\)& 5.3        &  0.04      &438        &  0.120(4)   &  0.037    & 55  & 1.19  & 1.08  \\
d&    \(12^3 \times 24\)& 5.3        &  0.05      &302        &  0.129(1)   &  0.050    & 40  & 1.17  & 1.10  \\
\hline
\end{tabular}
\end{center}
\caption{Parameters for the simulations; the first column denotes the
run,  for later reference. The gauge coupling is $\beta_1$, the bare quark
mass parameter $a m$, HMC-time denotes the  length of the run (number of
trajectories), and the lattice spacing has been determined via the Sommer
parameter. In the last two columns we give the finite volume correction
parameter $\rho$ and the renormalization factor for the condensate
obtained in a quenched setting \cite{GaGoHu04}, as discussed in Sect.
\ref{sec:results}. \label{tab:runparameters}}
\end{table}

In the table we also give the number of configurations of the run sequences
where the low lying spectrum of the Dirac operator has been extracted. 
Typically this was done every 5th configuration in order to  reduce
autocorrelation effects. Simulations with dynamical fermions are very costly, even more so for the
GW-type fermionic actions. For the moment we therefore have to
rely on comparatively small samples and lattices.

\subsection{Random Matrix Theory}\label{subsec:RMT}

Random Matrix Theory (RMT) was introduced into physics by Wigner \cite{Wi55}
who used it to  describe the apparently universal distribution of nuclear
level spacings. 
The connection between RMT and the low  lying spectrum
of the Dirac operator was made in \cite{ShVe93,Ve94a}.  Since then several 
lattice studies (mostly for the quenched situation) \cite{rmtlat} have 
confirmed the predictions 
of RMT. (See \cite{VeWe00,Da02,Ve05} for recent reviews in that context.)

The present understanding is that the generating  function for universal
features of the low lying Dirac spectrum is determined by chiral symmetry.
The microscopic spectrum thus follows the spectral properties of a random
matrix with symmetry like the Dirac operator, i.e., the so-called chiral 
Gaussian Unitary Ensemble (chGUE). This agreement is expected to hold in a
regime where the pion wave length is larger than the volume considered, such
that only pions dominate the dynamics and all other heavy degrees of 
freedom become irrelevant. In chiral
perturbation theory this is the so-called the $\epsilon$-regime
\cite{GaLe88}.  
Our pion mass is clearly too large for that with $m_\pi\approx 3/L$;
however, the smallest values of Im$(\lambda)$ leading to RMT
distributions have values such that
$\sqrt{2 |\textrm{Im}(\lambda) \Sigma|}/f_\pi \approx 1/L$ and thus we
may hope that RMT is applicable.

The QCD parameters,  quark mass and condensate, enter the RMT distributions
in form of dimensionless scaling variables $\mu=\Sigma V m$ and
$\zeta=\Sigma V \lambda $, where $\lambda$ denotes eigenvalues of the matrix
(or the imaginary part of the Dirac operator eigenvalues, respectively), and
$V$ is the volume. The procedure for calculating the exact RMT distributions 
$\rho_k[\nu ](\zeta)$ of the $k$th largest
eigenvalue  for a given number of flavors and in a fixed topological sector
$\nu$ have been worked out and checked in \cite{DaNi01DaHeNi00aAkDa04}.
In this way RMT allows one to determine $\Sigma$ from
comparison of numerically computed eigenvalue distributions with the
universal distributions given in terms of the scaling variables. 
Since our operator is not exactly chiral, we define our masses through the 
Axial Ward Identity (the so-called AWI-mass)
to take into account possible additive renormalization. 

Even though the predictions of RMT are true only in the infinite volume 
limit with $m$ tending to zero keeping $\mu$ fixed, the distributions are
approached already for comparatively small systems 
(for recent applications cf. \cite{DeLiSc06,HaHo06}).
The obtained value of $\Sigma$, although corresponding to the parameter of
infinitely sized random matrices, is not yet the condensate in a lattice
simulation. There the geometry (hypercubic, different space- and time
extent) provides extra correction factors which may be computed from sub-leading
chiral perturbation theory. One also has to consider scale dependent
renormalization factors. We discuss both in Sect. \ref{sec:results}.

As emphasized, e.g., in \cite{DeLiSc06}, the bias introduced by histogramming the data
may be avoided using the Kolmogorov-Smirnov (KS) test (cf. \cite{PrFlTe99}
for details). To apply the KS test one first creates a cumulative distribution
function of both the data as well as the probability function from which it is
supposed to be  drawn and compares the two.  We will use it to test whether the
RMT distribution, for fixed $\nu$ and fermion mass, for some value of $\Sigma$ describes the distribution of the
individual eigenvalues of the CI Dirac operator.  Given a data set, we find the
best fit to the  RMT distribution with the single parameter ``$\Sigma^{1/3}$"
(which minimizes the distance $D$ between the data cumulative distribution and
the RMT expectation) and quote the probability we obtain for  that value of
$\Sigma^{1/3}$.

\section{Results: Condensate}\label{sec:results}

For exact GW-operators the eigenvalues are exactly on the GW-circle, which
may be projected to the imaginary axis for further analysis, as done in
\cite{DeLiSc06}. In our case the eigenvalues typically are close to, but not
exactly on the circle. For a sample of the localization see Ref.
\cite{LaMaOr06}. We therefore use the distribution of the
values Im$(\lambda)$  for comparing with the
RMT cumulative distribution. The approximation introduces a systematic
error. This adds to the distortion effects due to lattice shape and finite
volume and thus we cannot expect our data to approach perfect RMT
distributions. We expect, however, that the central scaling properties of
the peak position survive in good approximation.

For the mass parameter of the RMT distributions we use the AWI-mass as 
determined in \cite{LaMaOr05c}.
Only real eigenmodes carry chirality and these are used to determine the
topological sector $\nu=n_--n_+$, where $n_\pm$ denotes the number of real
modes with right or left chirality. In most (97\%) configurations all the
real modes have the same chirality.

\begin{figure}[t]        
\includegraphics[width=\textwidth,clip]{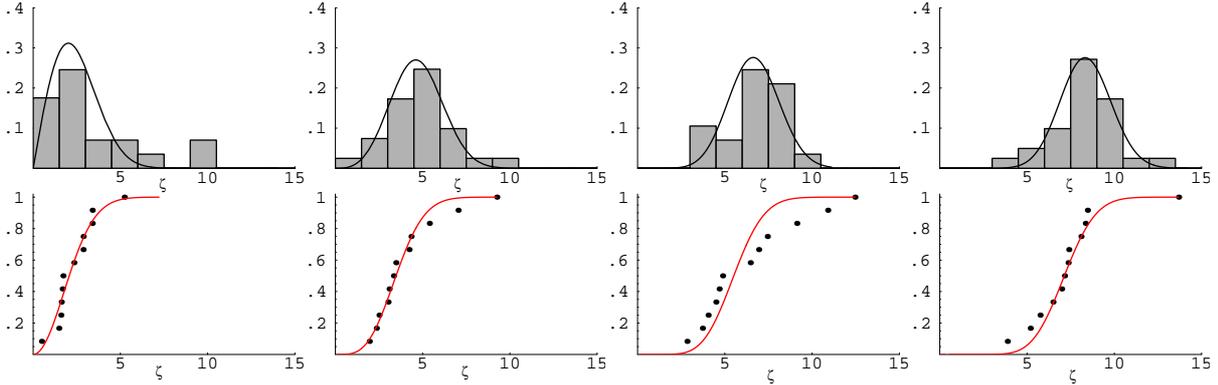}
\caption{Normalized histograms and cumulative distributions for run (a).
The corresponding probabilities ($Q_{KS}(k,\,\nu )$) are :  
 $Q_{KS}(1,\,0)$=0.50, 
 $Q_{KS}(1,\,1)$=0.99, 
 $Q_{KS}(2,\,0)$=0.90, 
 $Q_{KS}(2,\,1)$=0.96.
The abscissa $\zeta=\Sigma V \lambda $,
where $\lambda$ denotes the imaginary part of the corresponding 
eigenvalues.
}
\label{fig:runa}
\end{figure}

\begin{figure}[t]
\includegraphics[width=\textwidth,clip]{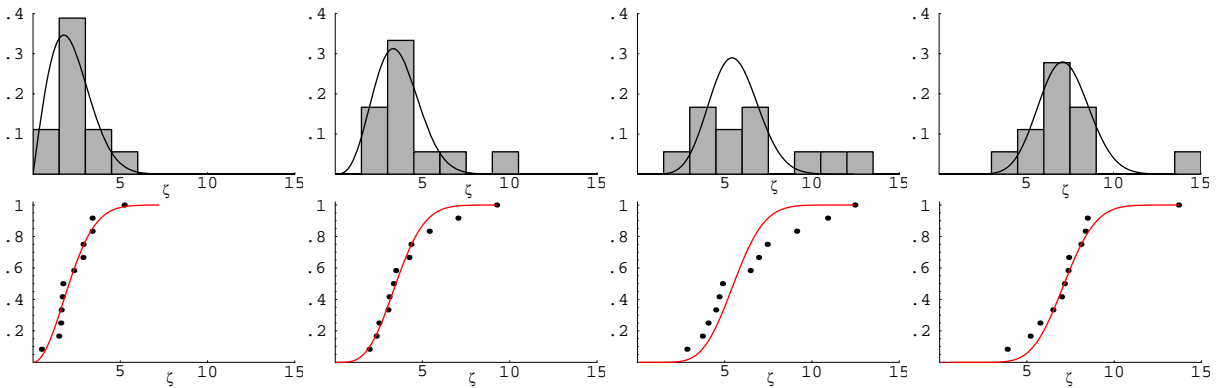}
\caption{Like Fig.\ \ref{fig:runa}, but now for run (d):  
 $Q_{KS}(1,\,0)$=0.98, 
 $Q_{KS}(1,\,1)$=0.99, 
 $Q_{KS}(2,\,0)$=0.74, 
 $Q_{KS}(2,\,1)$=0.99.
}
\label{fig:rund}
\end{figure}

In Figs.\ \ref{fig:runa} and \ref{fig:rund} we show some of the histograms
and cumulative distributions together with the theoretical cumulative
distributions for the  optimal value of $\Sigma$.

The results for all cumulants that have been analyzed are given in Table
\ref{tab:sigmaresults}. By $\Sigma^{1/3}$ we denote the values extracted
from the distributions according to the RMT formulas.  The given values
maximize the KS-probability for the distribution. The discrepancy between
the sum of the number of values entering the  histograms and the total
number of configurations analyzed is due to the configuration in higher
topological sectors.

The statistical errors for $\Sigma^{1/3}$ given in the table have been
calculated by using statistical bootstrap on the data. From each set of data
for a given run, $k$, and $\nu$ we produce several sets which are reanalyzed
to  give the spread in the resulting condensate. For each sector we have
just few eigenvalues and therefore the statistical fluctuation is
significant and may be even underestimated. 

We choose to fit the distributions for each $k$, $\nu$ set
individually. A joint fit would somewhat obscure the qualitative differences
which we did observe.
Naively, RMT is expected to give a unique $\Sigma$  for all these samples
for large enough statistics. However, as mentioned,
we cannot expect exact RMT
distribution shapes even for large enough statistics. 
In view of these limitations the observed differences of values from the
individual fits given in Table\ \ref{tab:sigmaresults} are not surprising. 
From the variation of the values we may derive an estimate for a
systematic error  (which does not cover all possible sources like scaling
with the lattices spacing  and properties of the Dirac operator).

\begin{table}
\begin{center}
\begin{tabular}{lrrclrr}
\hline
run    &  $k$  &     $\nu$ &\# values& $-\Sigma^{1/3}$[MeV]  & $Q_{KS}$ & $D_{\rm obs}$\\
\hline
(a)&     1  &     0&  19   & 325      (  25) &     0.50 &	0.18\\
   &     1  &     1&  27   & 280     (  5 ) &      0.99 &	0.08\\
   &     2  &     0&  19   & 288     (  3 )  &     0.93 &	0.12\\
   &     2  &     1&  27   & 284     (  3 ) &      0.97 &	0.09\\
\hline
   & \multicolumn{3}{l}{average:}&
\multicolumn{3}{l}{285 (3)(21)}\\
\hline
(b)&     1  &     0&  8  & 279       (  10) &     1.00 &	0.08\\
   &     1  &     1&  20 & 338       (  11) &     1.00 &	0.08\\
   &     2  &     0&  8  & 290       (  6)  &     0.77 &	0.22\\
   &     2  &     1&  20 & 313       (  9)  &     0.78 &	0.14\\
\hline
   & \multicolumn{3}{l}{average:}&
\multicolumn{3}{l}{301 (20)(23)}\\
\hline
(c)&     1  &     0&  16  & 291      (  20 )&       0.62 &	0.18\\
   &     1  &     1&  21  & 305      (  9 )&       0.87 &	0.13\\
   &     2  &     0&  16  & 289      (  6 )&        0.99 &	0.11\\
   &     2  &     1&  21  & 301      (  7 )&        0.29 &	0.21\\
\hline
   & \multicolumn{3}{l}{average:}&
\multicolumn{3}{l}{294 (7)(7)}\\
\hline
(d)&     1  &     0&  12  & 259     (  4 ) &      0.98 &	0.13\\
   &     1  &     1&  12  & 300     (  6 ) &      1.00 &	0.09\\
   &     2  &     0&  12  & 286     (  4 ) &      0.74 &	0.19\\
   &     2  &     1&  12  & 294     (  2 ) &      1.00 &	0.09\\
\hline
   & \multicolumn{3}{l}{average:}&
\multicolumn{2}{l}{288 (13)(16)}\\
\hline
\end{tabular}
\end{center}
\caption{Results for the optimal value of $\Sigma$ obtained from fitting the
individual cumulants to the RMT distributions, adjusting the parameter
$\Sigma^{1/3}$. The runs are ordered as in Table \ref{tab:runparameters}. In
the table $k$ denotes the $k^{th}$ zero in the topological sector $\nu$ (0
or 1). The number of eigenvalues entering the individual distributions is
given (\# values), as well as the KS-probability and the best value of the
distance $D_{\rm obs}$. \label{tab:sigmaresults}}
\end{table}

The sensitivity of the KS tests for a cumulative distribution function 
$P(x)$ is not independent  of $x$. In fact the KS test tends to be most
sensitive around the median value $P(x)=0.5$ and less sensitive at the
extreme ends of the distribution where $P(x)$ is near 0 or 1. The result is
that while KS tests are usually good for finding shifts in probability
distribution, they are  not so good at finding spreads which generally
affect the tails more than the median. Physically, in our case, this means
that a high confidence level in the KS tests indicates that the position  of
the maximum in the eigenvalue distribution is well predicted by RMT, but
does not tell us much  about the width of the distributions.

In order to give an overall estimate we average our central values for
$\Sigma^{1/3}$ for the distributions for each run, weighted by the
KS-probability and the inverse statistical error squared, leading to the
numbers given in the summary lines of Table\ \ref{tab:sigmaresults}. A
precise definition of the statistical error is hardly possible due to the
interplay of different probability factors. We estimate the statistical
error from the variance around the central value, again respecting the
individual statistical errors and the KS-probabilities. The first error
following the mean value denotes the statistical one and the second provides
an estimate of the systematic  error as discussed.

For a check of the quality of our error estimates we also compute the best
fit values  of $\Sigma^{1/3}$ for k=3, $\nu$=0 and see if they lie within
the error bars of the mean values for each lattice as given in Table
\ref{tab:sigmaresults}. For the runs  $a,b,c,d$ we obtain for $\Sigma^{1/3}$, the 
values 290(4), 290(2), 288(13), 281(4) with $Q_{KS}$ of  0.98, 0.76, 0.13, 0.48
respectively. All these values are   consistent within statistical errors
with the mean values obtained from the fits  to k=1 ,2 and $\nu$=0, 1. 

Leading order chiral perturbation theory provides corrections to the
condensate $\Sigma$ depending of the lattice  shape and the volume. The
geometry (hypercubic, different space- and time extent) provides extra
correction factors which have been discussed in sub-leading chiral
perturbation theory. The condensate $\Sigma(V)$ as  obtained from the finite
volume simulation has to be divided by  a correction factor $\rho$ given by
\cite{HaLe90,Da02}
\begin{equation}
\Sigma(\infty)=\Sigma(V) / \rho \quad\textrm{with} \quad
\rho = 1 + \frac{N_f^2-1}{N_f}\frac{\beta}{f_{\pi}^2 L^2}\;.
\end{equation}
Here $L=V^{1/d}$, $f_{\pi}$ is the pion decay constant,  and $\beta$
depends upon the lattice geometry. The procedure for  calculating this
coefficient in general is outlined in \cite{HaLe90}.  Here we only quote 
the result relevant for our geometry, i.e., for a lattice twice as long in
time  direction than in the other three: $\beta = 0.0836011$.

The values of $f_\pi$ for runs (a)--(d) have been measured
\cite{LaMaOr05c}.  The extrapolation to the chiral limit is compatible with
the physical value 93\ MeV, within a 10\% error margin. Since the correction
factor applies to the chiral limit, we can estimate it using the physical
value for the decay constant and the lattice spacings as determined in
\cite{LaMaOr05c}. We give the factors in Table \ref{tab:runparameters}.

One also has to consider scale dependent renormalization factors $Z_S=1/Z_m$
in order to relate to, e.g., the \ensuremath{\overline{\textrm{MS}}}  scheme
values, $Z^{(r)}=Z_S \,\Sigma$. For the quenched case these have been
determined in \cite{GaGoHu04} for various lattice spacings, in the
chiral limit, leading to
values ranging from 1.13 (at $a=0.148$ fm) to 0.96  (at $a=0.078$ fm). We
quote the factors $Z_{S,q}$ obtained from an interpolation of these quenched
values in Table\ \ref{tab:runparameters}.  For the lattice spacings studied
here they are all close to 1.1. We have not determined them for the dynamical
simulation (where an extrapolation to the chiral limit would not be  very
stable) but expect similar values. 

Table \ref{tab:sigmaresults} gives the uncorrected mean values for the condensate.
In order to compare with the continuum \ensuremath{\overline{\textrm{MS}}} 
values we therefore multiply the $\Sigma$ values of the table
with $(Z_S/\rho)$.
For the runs (a)--(d) this results in final values
\begin{eqnarray}
\Sigma^{(r)}(a)&=&(-274(3)(20)\textrm{~MeV})^3\,,\nonumber\\
\Sigma^{(r)}(b)&=&(-293(20)(23)\textrm{~MeV})^3\,,\nonumber\\
\Sigma^{(r)}(c)&=&(-285(7)(7)\textrm{~MeV})^3\,,\nonumber\\
\Sigma^{(r)}(d)&=&(-282(13)(16)\textrm{~MeV})^3\,.
\end{eqnarray}
The weighted average for the mean, with a simple
average for the errors gives an overall conservative
estimate of $\Sigma =(-276(11)(16)\textrm{~MeV})^3$. This is compatible with
determinations  by other groups and also with our own determination based on
the GMOR relation. 

\section{Discussion and Conclusion}

In this article we have extracted the quark condensate by comparing the low
lying  eigenvalue spectra of CI-Dirac operator with Random Matrix Theory.
The advantage of  this method is that it gives the condensate directly in
the chiral limit already on moderate sized lattices. By maximizing the 
probability in the Kolmogorov-Smirnov tests we obtained the chiral
condensate for  each of the $k (= 1,\,2)$ eigenvalue distributions in
fixed topological sectors  $\nu (= 0,\,1)$. Our final estimate after
correcting for renormalization and  finite volume effects is
$\langle\bar\psi\psi\rangle =(-276(11)(16)\textrm{~MeV})^3$  which is
consistent with determinations by other authors. Random Matrix Theory seems
to predict the peaks of the individual distributions  quite well. The widths
of the distributions on the other hand are not so well  matched.

\section{Acknowledgments}

We want to thank Poul Damgaard, Tom DeGrand, Christof Gattringer, Anna
Hasenfratz, Stefan Schaefer, Kim Splittorff, and Peter Weisz for helpful
discussions and comments. Support by Fonds zur F\"orderung der 
Wissenschaftlichen Forschung in \"Osterreich (FWF projects P16310-N08 and DK
W1203-N08) is gratefully acknowledged. The calculation have been done on the
Hitachi SR8000 at the Leibniz Rechenzentrum in Munich and at the Sun Fire
V20z cluster of the computer center of Karl-Franzens-Universit\"at, Graz,
and we want to thank both institutions for support.



\begin{thebibliography}{10}

\bibitem{GeOaRe68}
M.~Gell-Mann, R.~J. Oakes, and B.~Renner,
\newblock Phys. Rev. {\bf 175}, 2195 (1968).

\bibitem{We79GaLe84}
S.~Weinberg,
\newblock Physica A {\bf 96}, 327 (1979).
J.~Gasser and H.~Leutwyler,
\newblock Ann. Phys. {\bf 158}, 142 (1984).

\bibitem{Na89Ja02}
S.~Narison,
\newblock Phys. Lett. B {\bf 216}, 191 (1989).
M.~Jamin,
\newblock Phys. Lett. B {\bf 538}, 71 (2002), hep-ph/0201174.

\bibitem{ArShVe0406}
A. Armoni, M. Shifman, and G. Veneziano,
\newblock Phys. Lett. B {\bf 579}, 384 (2004), hep-th/0309013.
A. Armoni, G. Shore, and G. Veneziano,
\newblock Nucl. Phys. B {\bf 740} 23 (2006), hep-th/0511143.

\bibitem{GaHuLa05a}
C.~Gattringer, P.~Huber, and C.~B. Lang,
\newblock Phys. Rev. D {\bf 72}, 094510 (2005), hep-lat/0509003.

\bibitem{GuBh97}
R.~Gupta and T.~Bhattacharya,
\newblock Phys. Rev. D {\bf 55}, 7203 (1997), hep-lat/9605039.

\bibitem{Mc05}
C.~McNeile,
\newblock Phys. Lett. B {\bf 619}, 124 (2005), hep-lat/0504006.

\bibitem{AoBlCh05}
Y.~Aoki {\em et~al.},
\newblock Phys. Rev. D {\bf 72}, 114505 (2005), hep-lat/0411006.

\bibitem{DeHoLi06}
T.~DeGrand, R.~Hoffmann, Z.~Liu, and S.~Schaefer,
\newblock Phys. Rev. D {\bf 74}, 054501 (2006), hep-th/0605147.

\bibitem{DeLiSc06}
T.~DeGrand, Z.~Liu, and S.~Schaefer,
\newblock Phys. Rev. D {\bf 74}, 094504 (2006), hep-lat/0608019.

\bibitem{HaHo06}
A.~Hasenfratz and R.~Hoffmann,
\newblock Phys. Rev. D {\bf 74} 114509 (2006), hep-lat/0609067. 
\newblock PoS {\bf LAT2006}, 210 (2006), hep-lat/0609070.

\bibitem{GiWi82}
P.~H. Ginsparg and K.~G. Wilson,
\newblock Phys. Rev. D {\bf 25}, 2649 (1982).

\bibitem{Ga01aGaHiLa00}
C.~Gattringer,
\newblock Phys. Rev. D {\bf 63}, 114501 (2001), hep-lat/0003005.
C.~Gattringer, I.~Hip, and C.~B. Lang,
\newblock Nucl. Phys. {\bf B597}, 451 (2001), hep-lat/0007042.

\bibitem{GaGoHa03a}
C.~Gattringer {\em et~al.},
\newblock Nucl. Phys. {\bf B677}, 3 (2004), hep-lat/0307013.

\bibitem{GaGoHu04}
C.~Gattringer, M.~G{\"o}ckeler, P.~Huber, and C.~B. Lang,
\newblock Nucl. Phys. {\bf B694}, 170 (2004), hep-lat/0404006.

\bibitem{LaMaOr05ab}
C.~B. Lang, P.~Majumdar, and W.~Ortner,
\newblock Proc. Sci. {\bf LAT2005}, 124 (2005), hep-lat/0509004.
C.~B. Lang, P.~Majumdar, and W.~Ortner,
\newblock ibid., 131 (2005), hep-lat/0509005.

\bibitem{LaMaOr05c}
C.~B. Lang, P.~Majumdar, and W.~Ortner,
\newblock Phys. Rev. D {\bf 73}, 034507 (2005), hep-lat/0512014.

\bibitem{LaMaOr06}
C.~B. Lang, P.~Majumdar, and W.~Ortner,
\newblock {D}irac eigenmodes in an environment of dynamical fermions,
\newblock in {\em Sense of beauty in physics}, Ed. M. D'Elia et al.
(Pisa University Press: 2006), hep-lat/0512045.

\bibitem{LuWe85}
M.~L{\"u}scher and P.~Weisz,
\newblock Commun. Math. Phys. {\bf 97}, 59 (1985).

\bibitem{MoPe04}
C.~Morningstar and M.~Peardon,
\newblock Phys. Rev. D {\bf 69}, 054501 (2004), hep-lat/0311018.

\bibitem{Wi55}
E.~P. Wigner,
\newblock Ann. Math. {\bf 62}, 548 (1955).

\bibitem{ShVe93}
E.~V. Shuryak and J.~J.~M. Verbaarschot,
\newblock Nucl. Phys. A {\bf 560}, 306 (1993), hep-th/9212088.

\bibitem{Ve94a}
J.~J.~M. Verbaarschot,
\newblock Phys. Rev. Lett. {\bf 72}, 2531 (1994).

\bibitem{rmtlat}
M.~Berbenni-Bitsch {\em et al.}, Phys. Rev. Lett. {\bf 80} 1146 (1998),
hep-lat/9704018, 
\newblock R.~Edwards, U.~Heller and R.~Narayanan, Phys. Rev. D {\bf 60}, 034502 (1999),
hep-lat/9901015,
\newblock M.~Gockeler {\em et al.}, Phys. Rev. D {\bf 59}, 094503 (1999),
hep-lat/9811018,
\newblock L.~Giusti, M.~L\"uscher, P.~Weisz and H.~Wittig, JHEP {\bf 0311}, 023 (2003),
hep-lat/0309189,
\newblock R.~Narayanan and H.~Neuberger, Nucl. Phys. B {\bf 696}, 107 (2004),
hep-lat/0405025,
\newblock E.~Follana, A.~Hart and C.T.H.~Davies, PoS {\bf LAT2005}, 298 (2006),
hep-lat/0509177.

\bibitem{VeWe00}
J.~J.~M. Verbaarschot and T.~Wettig,
\newblock Ann. Rev. Nucl. Part. Sci. {\bf 50}, 343 (2000), hep-ph/0003017.

\bibitem{Da02}
P.~H. Damgaard,
\newblock Nucl. Phys. B (Proc. Suppl.) {\bf 106}, 29 (2002), hep-lat/0110192.

\bibitem{Ve05}
J.~J.~M. Verbaarschot,
\newblock {QCD}, {C}hiral random matrix theory and integrability,
Les Houches Summer School on Applications of Random Matrices in Physics 
(2004), hep-th/0502029.

\bibitem{GaLe88}
J.~Gasser and H.~Leutwyler,
\newblock Nucl. Phys. B {\bf 307}, 763 (1988).

\bibitem{DaNi01DaHeNi00aAkDa04}
P.~H. Damgaard and S.~M. Nishigaki,
\newblock Phys. Rev. D {\bf 63}, 045012 (2001), hep-th/0006111.
P.~H. Damgaard, U.~M. Heller, R.~Niclasen, and K.~Rummukainen,
\newblock Phys. Lett. B {\bf 263}, 495 (2000), hep-lat/0007041.
G.~Akemann and P.~H. Damgaard,
\newblock Phys. Lett. B {\bf 583}, 199 (2004), hep-th/0311171.

\bibitem{PrFlTe99}
W.~H. Press, S.~A. Teukolsky, W.~T. Vetterling, and B.~P. Flannery,
\newblock {\em {N}umerical recipes in {C}}, 2nd ed. (Cambridge University
  Press, Cambridge New York, 1999).

\bibitem{HaLe90}
P.~Hasenfratz and H.~Leutwyler,
\newblock Nucl. Phys. B {\bf 343}, 241 (1990).

\bibitem{JLQCD}
H.~Fukaya {\em et al.}, hep-lat/0702003.

\end{thebibliography}
\end{document}